\documentclass[%
 reprint,
 floatfix,
 amsmath,
 amssymb,
 aps,
 prl,
]{revtex4-2}

\usepackage{xcolor}
\usepackage{graphicx}
\usepackage{dcolumn}
\usepackage{bm}
\usepackage{microtype}
\usepackage{physics} 
\usepackage[normalem]{ulem}

\usepackage{hyperref}

\begin{document}

\preprint{APS/123-QED}

\title{
Experimental particle production in time-dependent spacetimes: a one-dimensional scattering problem}

\author
{Marius Sparn,$^{1\ast}$ Elinor Kath,$^{1}$ Nikolas Liebster,$^{1}$ Jelte Duchene,$^{1}$ Christian F. Schmidt,$^{2}$ Mireia Tolosa-Simeón,$^{3}$ Álvaro Parra-López,$^{4}$ Stefan Floerchinger,$^{2}$ Helmut Strobel$^{1}$ and Markus K. Oberthaler$^{1}$ \\
\normalsize{$^{1}$Kirchhoff-Institut f\"{u}r Physik, Universit\"{a}t Heidelberg, \\
    Im Neuenheimer Feld 227, 69120 Heidelberg, Germany}\\
\normalsize{$^{2}$Theoretisch-Physikalisches Institut, Friedrich-Schiller-Universit\"{a}t Jena, \\
    Max-Wien-Platz 1, 07743 Jena, Germany}\\
\normalsize{$^{3}$Institut für Theoretische Physik III, Ruhr-Universität Bochum, Bochum, Germany}\\
\normalsize{$^{4}$Departamento de Física Teórica and IPARCOS, Facultad de Ciencias Físicas, Universidad Complutense de Madrid, Ciudad Universitaria, 28040 Madrid, Spain}\\
\normalsize{$^\ast$scatteringanalogy@matterwave.de}
}

\date{December 24, 2024}

\begin{abstract}
We experimentally study cosmological particle production in a two-dimensional Bose-Einstein condensate, whose density excitations map to an analog cosmology.
The expansion of spacetime is realized with tunable interactions.
The particle spectrum can be understood through an analogy to quantum mechanical scattering, in which the dynamics of the spacetime metric determine the shape of the scattering potential.
Hallmark scattering phenomena such as resonant forward scattering and Bragg reflection are connected to their cosmological counterparts, namely linearly expanding space and bouncing universes.
We compare our findings to a theoretical description that extends beyond the acoustic approximation, which enables us to apply the model to high-momentum excitations.
\end{abstract}

\maketitle
\textit{Introduction}--Cosmological particle production is a striking result of time-dependent spacetimes, where the dynamics can result in the generation of excitations of a quantum field even from an initial vacuum state.
This phenomenon is relevant in cosmological models \cite{Schroedinger1939,Parker1969,Mukhanov2007}, and has been the topic of a variety of studies in analog cosmology.
In Bose-Einstein-Condensates (BEC) \cite{Visser2001,Visser2003,Fedichev2004,Fischer2004,Gardiner2007,Weinfurtner2007,Weinfurtner2009,Carusotto2010,Prain2010,Jaskula2012,Hung2013,Eckel2018,Chen2021,Steinhauer2022,Banik2022,Viermann2022}, the mean-field density distribution defines an effective spacetime metric, while phononic density-fluctuations are analogous to a cosmological quantum field \cite{Mireia2022}. 
Since the speed of sound sets a local physical ruler, time-dependence of the metric can be implemented by adjusting the sound speed via the inter-atomic interaction.

In this Letter, we give a new perspective on the particle creation process, by reformulating particle-production in two spatial dimensions as reflection on a one-dimensional quantum mechanical scattering potential \cite{Schmidt2024}.
This work establishes an intuitive framework with which to understand the otherwise convoluted and out of equilibrium process of particle production.

\begin{figure}[b]
     \centering
     \includegraphics{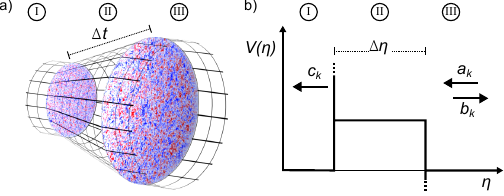} 
     \caption{Particle production and its scattering analog. 
     (a) A quantum field on a time-dependent spacetime develops excitations, a process known as cosmological particle production. 
     In a 2D Bose gas of fixed size the  expansion is realized by changing the scattering length. 
     Field excitations manifest in increased density fluctuations, as visible in the shown density contrasts.
     (b) Predicting the particle spectrum is equivalent to solving a scattering problem in one dimension, where the time dependence of the scale factor $a(t)$ sets the potential $V(\eta)$ in conformal time $\eta$.
     Non-vanishing reflected amplitudes $b_k$ correspond to the production of particles in a momentum mode $k$.
     }
     \label{Fig1}
\end{figure}

\textit{Scattering analogy}--As has been discussed \cite{Visser2003,Gardiner2007,Mireia2022}, phononic excitations of a quasi two-dimensional BEC with repulsive interactions define a massless, relativistic, scalar field $\phi$, which manifest in the form of density fluctuations \cite{Supplementary}.
For homogeneous density distributions, the field is governed by a spatially flat Friedmann-Lemaître-Robertson-Walker metric,
\begin{equation}
    \mathrm{d}s^2 = - \mathrm{d}t^2 + a^2(t)\left(\mathrm{d}r^2 + r^2 \mathrm{d}\varphi^2\right),
    \label{eq:metric}
\end{equation}
in polar coordinates $r$ and $\varphi$. 
The scale factor $a(t)$ relates to the speed of sound $c_s(t)$ in the condensate as $a(t) = 1/c_s(t)$, where $c_s$ is given by the linear low momentum limit of the Bogoliubov dispersion relation. 
The field $\phi$ can be expanded in time-dependent mode functions $v_k(t)$ \cite{Supplementary}, and the time evolution of each $v_k$ is described by
\begin{equation}
    \ddot{v}_k + 2\frac{\dot a}{a}\dot{v}_k+\frac{k^2}{a^2}v_k = 0,
    \label{eq:mode-equation_lab}
\end{equation}
which is a Klein-Gordon type equation with an extra Hubble-friction term in form of a first derivative \cite{Birrell1982,Mukhanov2007}.
In a static spacetime ($\dot a=0$), this equation is solved by wave solutions with positive and negative frequencies, and occupation numbers are constant and well defined.
Dynamics of $a(t)$ linking two static regions, however, give rise to a Hubble-friction term, breaking time-reversal symmetry and energy conservation. This can result in the generation of particle pair excitations of the field. A field without excitations initially, i.e. in the vacuum state, is transformed into a two-mode squeezed state after a change of the scale factor.

The Hubble-friction term can be absorbed by performing a coordinate transformation to conformal time $\mathrm{d}\eta = \frac{1}{a(t)} \mathrm{d}t = c_s(t) \mathrm{d}t$ and considering the re-scaled mode-functions $\psi_k(\eta) = \sqrt{a(\eta)}v_k(\eta)$.
With these transformations Eq. (\ref{eq:mode-equation_lab}) takes the form of a one-dimensional (non-relativistic) Schrödinger type equation in conformal time
\begin{equation}
    \left(-\frac{\mathrm{d}^2}{\mathrm{d}\eta^2} + V(\eta)\right)\psi_k(\eta)  = k^2 \psi_k(\eta),
    \label{eq:mode-equation_SE}
\end{equation}
with the (scattering) potential
\begin{equation}
V(\eta) = \frac{1}{4}\,\dot{a}^2(t(\eta)) + \frac{1}{2}\,\ddot{a}(t(\eta))\,a(t(\eta)),
\label{eq:Potential}
\end{equation}
where the dots denote the derivative with respect to laboratory time.
Within this description, the time dependent scale factor $a(t(\eta))$ translates into a scattering potential, and the particle production is connected to the momentum dependent scattering amplitudes (details in \cite{Schmidt2024}).

\begin{figure}
     \includegraphics{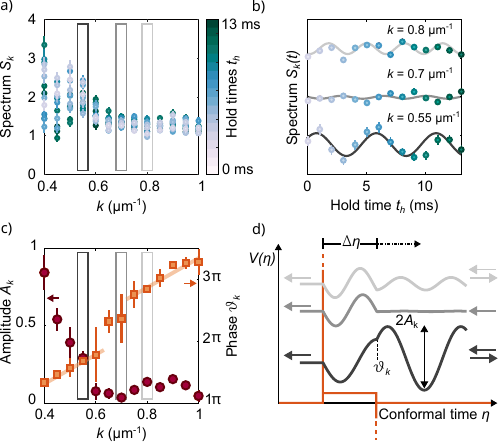}
     \caption{Systematic study of scattering states. 
     a) Density fluctuation power spectra $S_k$ calculated from a rescaled density contrast at different times $t_h$ after a linear expansion in the scale factor over $3\,$ms. 
     b) Observation of oscillating density fluctuations as expected from interference of the incoming and reflected waves (Sakharov oscillations). Solid lines are cosine fits.
     c) The extracted amplitude vanishes at a specific $k$, accompanied by a $\pi$ phase shift.
     Orange lines are a guide to the eye.
     d) Scattering potential (orange) and corresponding scattering states $|\psi_k|^2$ (gray) for the three indicated momenta. 
     The resonant $k$ shows no reflection, analogous to the Ramsauer-Townsend effect in electron scattering.
     Errors are $1\sigma$ standard errors of the mean (a,b) or from the fit (c).
     }
     \label{Fig2}
\end{figure}

Figure \ref{Fig1} shows a prototypical example of the analogy, using a linearly expanding spacetime (a). 
In regions I and III (where the scale factor is constant), the potential vanishes and the solutions take the form of waves (b). 
In region II, however, the dynamics of the scale factor $a(t) = a_\mathrm{min}(1 + H_0 t) $ result in a non-zero potential, in this case a box potential with height $H_0^2/4$ and additional peaks in the form of delta distributions from the abrupt start and end of the ramp.
An incoming wave from the right ($a_k \mathrm{e}^{-i k \eta}$) is scattered by the potential barrier, resulting in a transmitted ($c_k\mathrm{e}^{-ik \eta}$) and a reflected ($b_k\mathrm{e}^{i k \eta}$) part.
The transmitted wave corresponds to the initial vacuum state before the expansion in region I, and the reflected wave corresponds to negative frequency waves in region III.
This reflection on the potential barrier corresponds to particle production, which appears as enhanced density fluctuations. 

In region III incoming (positive frequency) and reflected (negative frequency) waves overlap, $\psi_k(\eta) = a_k e^{-\mathrm{i}k\eta} + b_k e^{\mathrm{i}k\eta}$.
This can be related to the experimentally accessible density-fluctuation power spectrum $S_k(\eta) = |\psi_k(\eta)|^2/\left(2|c_k|^2\right)$, with $|c_k|^2 = |a_k|^2 - |b_k|^2$.
Because the spacetime is static in region III, the interference of reflected and incoming waves changing in $\eta$ (Fig. \ref{Fig2}d)) translates into an oscillation of the power spectrum in the lab frame,
\begin{equation}
    S_k = 1/2 + N_k + \Delta N_k\cos(2\omega_k t_h+\vartheta_k),
    \label{eq:Sk}
\end{equation}
with $t_h$ the time in region III.
Here, $N_k = |b_k|^2/|c_k|^2$ is the occupation number of the mode, which corresponds to the produced particles.
$\Delta N_k=|a_kb_k|/|c_k|^2$ is the standing wave contribution of the scattering state, oscillating in time with twice their eigenfrequency $2\omega_k = 2c_sk$, and $\vartheta_k$ is the phase of this standing wave at the boundary between regions II and III ($t_h=0$).
An incoherent, \textit{e.g.}~thermal initial state can be captured by modifying the spectrum to $S_k\to(1+2N_k^\mathrm{in}) S_k$, where $N_k^\mathrm{in}$ is is the initial occupation in region I \cite{Supplementary,Mireia2022}. 
This captures stimulated particle production from initially occupied modes, and impacts all terms of $S_k$, such that the oscillation amplitudes measured in the experiment, $A_k$, differ from $\Delta N_k$ by this k-dependent factor.

\begin{figure}
     \centering
     \includegraphics{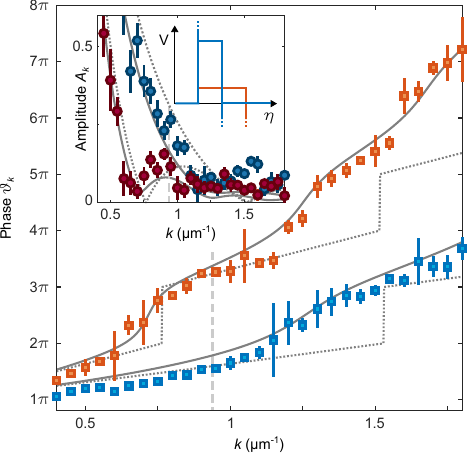}
     \caption{Comparison of theory and experiment in and beyond the acoustic approximation. Extracted amplitudes (inset) and phases after linear expansion by a factor of $\sqrt8$ over $\Delta t = 1.5\,\text{ms}$ (blue) and $\Delta t = 3.0\,\text{ms}$ (orange).
     Acoustic theory predictions (dotted) fit well in the regime $k<\xi^{-1}\sim 0.94\,\mu\text{m}^{-1}$, while a full Bogoliubov treatment (solid) gives quantitative agreement for all measured length scales.
     All errors are $1\sigma$ standard errors from the fit of the oscillations.}
     \label{fig:BoxPotential}
\end{figure}

\textit{Implementation}--The quantum field simulator is a BEC consisting of around $40,000$ atoms of $^{39}$K, tightly confined in the vertical direction using an optical lattice with $\omega_z = 2\pi\times1.5\,\text{kHz}$, resulting in an effectively two-dimensional system. 
Radial confinement is realized with a Digital-Micromirror Device for a configurable trap geometry.
Because we consider spatially flat spacetimes, we use homogeneous density distributions in a circular box trap.
Time dependence of the scale factor is implemented by globally modifying the speed of sound with the scattering length $a_s\propto c_s^2$ \cite{Viermann2022}, using a magnetic Feshbach resonance at $561\,\text{G}$ \cite{Etrych2023}.
Excitations forming at the walls of the box potential are minimized by adjusting the potential height during the ramp.
Two-dimensional density distributions are read out via high resolution absorption imaging at high magnetic fields \cite{Maurus2021}.
In order to extract the spectrum, two-point correlations of the density contrast are calculated and averaged over $\sim50$ shots, after discarding those that deviate more than $10\%$ from the mean atom number to ensure comparable sound speeds.
From these density-contrast correlations, the density fluctuation power spectrum is calculated via a Hankel-Transform, a transformation to momentum space that respects the radial symmetry of the correlations. 
To account for $k$-dependent imaging sensitivity, a  modulation transfer function is determined and corrected for in the final spectrum \cite{Hung2011}.
We measure the speed of sound $c_s$ independently via wave packet propagation.

\textit{Linear Expansion}--Figure\,\ref{Fig2} shows results for a linear expansion scenario.
This is realized by a ramp in scattering length from $400\,a_B$ to $50\,a_B$ of the form $a_{s} \propto 1/t^2$, which increases the scale factor $a$ linearly by a factor $\sqrt8$.
The spectrum is measured for different hold times $t_h$ in region III (Fig.\ref{Fig2}a). 
The oscillations described by Eq. \ref{eq:Sk} become apparent by plotting the value of the spectrum at a speciﬁc momentum $k$ as a function of time (Fig. \ref{Fig2}b).
We use cosine fits to extract oscillation amplitude $A_k$ and initial phase $\vartheta_k$, plotted in Fig. \ref{Fig2}c. 

A characteristic feature of this scenario is a minimum in oscillation amplitude at $k=0.7\,\mu\text{m}^{-1}$, accompanied by a rapid evolution in the extracted phase and large uncertainties; the otherwise linear evolution of the phase is shifted by $\pi$ between the left and right of the feature.
Exemplary oscillations shown in Figure \,\ref{Fig2}b demonstrate this feature: $k=0.55\,\mu\text{m}^{-1}$ and $k=0.8\,\mu\text{m}^{-1}$ show a difference in initial phase of roughly $\pi$.
This can be understood in the scattering framework, where the hold times in the lab frame are equivalent to measuring various points along the standing wave to the right of the potential as shown in Figure\,\ref{Fig2}d.
The box potential has a height of $\frac{1}{4}\dot a^2(\eta) = 0.01\,\mu\text{m}^{-2}$, indicating that momenta $k<0.1\,\mu\text{m}^{-1}$ would be classically reflected.
Quantum mechanically, higher momenta will still be reflected on the potential, which means their scattering states will have a standing wave contribution in region III. 
The exception are states with wavelengths that are resonant with the potential width $\Delta\eta$. 
Momenta $k \approx j\pi/\Delta\eta$, where $j$ is an integer, are not reflected but resonantly forward scattered instead. 
They thus appear as a minimum in the oscillation amplitude with ill-defined phase and correspond to a zero-crossing of the interference term. 
Momenta left and right of the resonance show opposite signs in the interference term, and therefore result in a shift of the phase over the resonance.
This is analogous to the Ramsauer-Townsend effect in electron scattering, where the scattering amplitude is drastically reduced as a function of momentum \cite{Ramsauer1921,Townsend1921}.

\begin{figure*}
     \centering
     \includegraphics{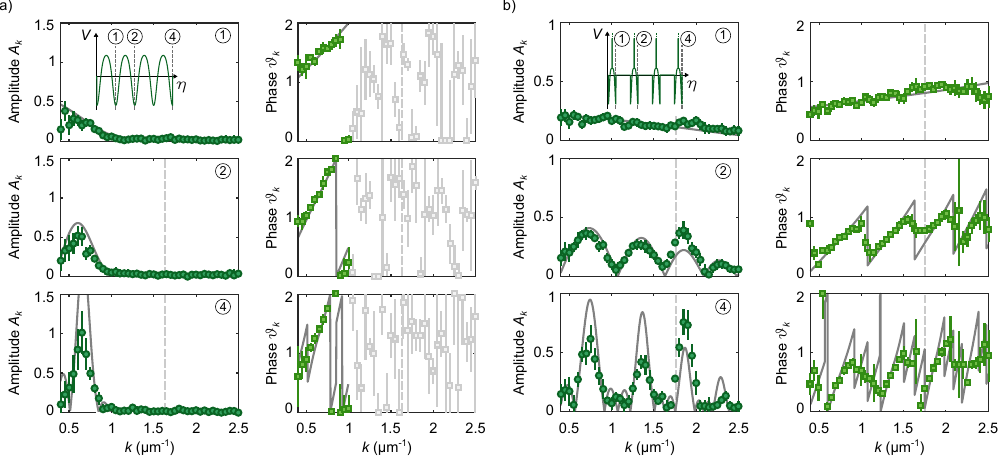}
     \caption{Bragg scattering in a periodically contracting and expanding spacetime. 
     a) A harmonically modulated scale factor results in a periodic potential (inset) dominated by a single Fourier component, leading to only one Bragg condition. Thus, a single peak at $\pi/\Delta \eta$ arises, accompanied by a slope in the phase (due to low amplitudes, phases at high momenta can not be reliably extracted, gray-shaded). 
     b) Realizing a series of sharp peaks in the potential landscape (inset) leads to higher Fourier components, and thus multiple Bragg conditions. 
     The experiments confirm the expected particle production at the corresponding length scales.
     In all graphs solid lines show Bogoliubov theory predictions and the corresponding $\xi^{-1}$ is indicated with dashed lines.
     All errors are $1\sigma$ standard errors from the fit of the oscillations.}
     \label{fig:Bragg}
 \end{figure*}
 
The scattering picture suggests that these resonances reoccur for higher multiples of $j$ and should shift with the width of the scattering barrier $\Delta\eta$. 
Figure \ref{fig:BoxPotential} shows a comparison between a $3\,$ms (orange) and a $1.5\,$ms (blue) ramp, which result in scattering barriers that differ by a factor two in width.
The two scenarios can be cleanly differentiated in their phases $\vartheta_k$.
In the phase $\vartheta_k$ of the previously discussed ramp we observe rapid phase evolution (i.e. resonant modes) at $k\sim 1.2\,\mu\text{m}^{-1}$ and $k\sim 1.6\,\mu\text{m}^{-1}$, in addition to the first resonance at $k\sim 0.7\,\mu\text{m}^{-1}$. 
Each of these resonances is accompanied by diminished oscillation amplitudes.
For the faster ramp, the resulting potential barrier is half as wide, and we therefore expect every second resonance of the slow ramp.
Indeed, we observe the first resonance at $k\sim 1.2\,\mu\text{m}^{-1}$. 
The precise position of the resonance is also influenced by the potential height, but this effect is too small to be resolvable.
Modes with momenta $k<0.2\,\mu\text{m}^{-1}$ would be tunneling modes that correspond to the decaying modes in cosmology \cite{Mukhanov2007,Weinberg2008}, but they are below the experimentally accessible range.

The theory prediction for the acoustic approximation is shown as the dotted line in Fig. \ref{fig:BoxPotential}.
While the experimental results agree with this prediction below the inverse healing length $\xi^{-1} = \sqrt2mc_s/\hbar$ (dashed line), for larger momenta the extracted phase deviates, because the dispersion significantly deviates from the linear approximation.
This is a common challenge in analog cosmology, and has to be handled for each experimental platform individually \cite{Robertson2016,Jacquet2020,Isoard2020}. Similarly, modified dispersion relations have also been discussed in the context of inflationary models for cosmology \cite{Niemeyer2001}.

We include the full Bogoliubov dispersion relation, $\omega_k = c_sk\sqrt{1 + \frac{1}{2}k^2\xi^2}$ \cite{Bogolyubov1947}, leading to a $k$-dependent metric and in turn potential \cite{Supplementary,Weinfurtner2009,Schmidt2024}.
Because the $k$-dependence changes during expansion, the symmetry of the box potential is broken, and the resonance condition is weakened.
As a result, the resonances do not have zero amplitude (see inset) and the phase feature is broadened, as can be seen by comparing the acoustic prediction to the full Bogoliubov description (solid line).
Because modes are forward scattered if they have multiples of $\pi$-phase evolution in the potential, the resonance positions shift to lower momenta as the dispersion relation deviates from the linear approximation.

In order to account for deviations in the precise shape of the experimental ramp, which lead to an effective difference in conformal time during the dynamics, we adjust both the linear and Bogoliubov models to include ramps that are $10\,\%$ shorter than programmed, and a final speed of sound $\sim 10\,\%$ lower than measured. 
All observed amplitudes $A_k$ are a factor two lower than the theoretical predictions, which have been scaled accordingly throughout the manuscript.

\textit{Periodic spacetime}--If the spacetime is subject to periodic expansions and contractions, the corresponding potential will also be periodic. 
Periodic potentials will result in a band structure with band gaps, where waves of specific energy and momentum can not propagate and therefore will be strongly reflected, known as the Bragg condition \cite{Bragg1913}.

The simplest periodic cosmology is a smooth transition into an oscillating scale factor, 
$a(t) = \frac{a_{\text{i}}+a_{\min}}{2}+\frac{a_{\text{i}}-a_{\min}}{2}\cos(\omega t)$,
which results in a sinusoidal scattering potential in the limit of small amplitude (compare equation \ref{eq:Potential}).
To implement this cosmology in the experiment, we periodically modulate the scattering length with the corresponding form starting from $200\,a_B$ to a maximum value of $400\,a_B$ with frequency $\omega/(2\pi) = 500\,$Hz.

Figure \ref{fig:Bragg} a) shows results for 1, 2, and 4 drive periods.
We find large amplitudes in one resonant momentum mode $k_{\text{res}}\sim 0.65\,\mu\text{m}^{-1}$. 
This corresponds to high reflection at $k_{\text{res}} = \omega/(2c_s)$, the edge of the first Brillouin zone, where the first band gap is expected.
The amplitudes grow with the number of oscillation periods, while the corresponding phases show a steepening in the slope around the peak.
The growth and sharpening of the peak can be understood as the onset of band structure. As the periodic nature of the potential becomes sharply defined the reflection increases.
Because this potential has no substantial contributions of higher harmonics, higher band gaps beyond the first are small and we find vanishing amplitudes.

In contrast, an anharmonic potential will result in band gaps at higher harmonics of the periodicity of the potential (Fig. \ref{fig:Bragg}b).
We produce such a potential with a cusp-like modulation of the scale factor. 
A single cusp is realised by a shape of $a(t) = \frac{a_{\text{i}}}{\sqrt2}\left[1-1/2|\cos(\omega t)|\right]^{-1/2}$ over half a period.
In between the cusps the scale factor is held constant at $a_{\text{i}}$ for $\Delta t = 2\text{ms}$, which sets the periodicity of the potential.
In the experiment the cusps reach from $200\,a_B$ to $400\,a_B$, and $\omega/(2\pi) = 800\,$Hz. 
The sudden switch in the slope results in peaks in the potential (see inset Fig. \ref{fig:Bragg}b).
In contrast to the harmonic potential, a single peak produces excitations over a broad range of momenta. 
For more than one period we find a series of resonances equally spaced in energy, each corresponding to the higher band gaps, and in good agreement with the prediction including the full Bogoliubov dispersion (solid line, see \cite{Supplementary}).
Each peak is accompanied by a slope in the phase and a smooth phase evolution between neighbouring peaks, resembling a smoothed out version of the theoretical prediction.
While imperfections in the experimentally implemented cusp shape are again corrected for by a $10\%$ adjustment in the speed of sound and a lag in the magnetic field control by $230\,\mu$s \cite{Supplementary}, these corrections are not necessary for the harmonic potential.

\textit{Conclusion}--We experimentally demonstrate the analogy between a scalar quantum field in time-dependent spacetimes and quantum mechanical scattering in a one-dimensional stationary Schrödinger equation.
By including Bogoliubov corrections in predictions obtained via scattering theory, we find good agreement to experimental data obtained from quantum field simulation in a BEC.
In addition to the discussed analogy, this establishes a framework to describe quasi-particle production in cold quantum gases.
For example, populating a single narrow band of momenta by sinusoidally modulating the scale factor instead of the scattering length could be employed to engineer far from equilibrium quantum states.

\section*{Acknowledgements} The authors thank Jean Dalibard for inspiring discussions. This work is supported by the Deutsche Forschungsgemeinschaft (DFG, German Research Foundation) under Germany’s Excellence Strategy EXC 2181/1 - 390900948 (the Heidelberg STRUCTURES Excellence Cluster), under SFB 1225 ISOQUANT - 273811115. This project was funded within the QuantERA II Programme that has received funding from the European Union’s Horizon 2020 research and innovation programme under Grant Agreement No 101017733 as well as the DFG under project number 499183856. N.L. acknowledges support by the Studienstiftung des Deutschen Volkes. C.F.S acknowledges support by the Studienstiftung des Deutschen Volkes and the Deutsche Forschungsgemeinschaft (DFG) under Grant No 406116891 within the Research Training Group RTG 2522/1. Á. P.-L. acknowledges support through the MICIN (Ministerio de Ciencia, Innovación y Universidades, Spain) fellowship FPU20/05603 and the projects PID2020-118159GBC44, and PID2022-139841NB-I00, COST (European Cooperation in Science and Technology) Actions CA21106 and CA21136.

\section*{Supplementary}

\subsection{Extraction of the sound speed}
The essential scale in the experiment is set by the sound speed. We use values which are obtained experimentally via measuring the propagation of localized density perturbations. 
For this, we prepare a density wave packet by creating a hole in the BEC with repulsive laser light. When this light is switched off a circular wave front moves outwards. By taking images after different times we track the position of the wave packet, and then get the speed of sound by fitting a slope to these positions. The speed of sound sets the scale factor $a$, the chemical potential $\mu$ and together with the known interaction strength the mean background density $\bar{n}_0$.
The values relevant for the scenarios discussed in the main text are $c_\mathrm{s}(50 a_0) = (1.07 \pm 0.04 )\, \mu\mathrm{m}/\mathrm{ms}$ for the linear expansion,  $c_\mathrm{s}(200 a_0) = (2.01 \pm 0.03) \, \mu\mathrm{m}/\mathrm{ms}$ for the anharmonic cusp-scalefactor and $c_\mathrm{s}(200 a_0) = (1.87 \pm 0.06) \, \mu\mathrm{m}/\mathrm{ms}$ for the harmonically modulated scale-factor. As stated in the main text, deviations between the experimentally realized ramps and the theoretical shapes are accounted for by adjusting the speed of sound.

\subsection{Extraction of spectra}
Starting from the experimentally extracted density distributions $n(x,y)$, we first exclude runs where the total atom number differs more than $10\,\%$ from the mean value. 
A renormalized density contrast is calculated via
\begin{equation}
    \delta_c(x,y) = \left(\frac{\bar{N}}{N}n(x,y)-\bar{n}(x,y)\right)\,\frac{\bar{n}^{1/2}(x,y)}{\bar{n}_0^{3/2}}\text{ ,}
\end{equation}
where $\bar{n}(x,y)$ is the mean density distribution over all shots and $\bar{n}_0$ is its value in the center of the condensate averaged over a $20\,$ pixel diameter. The factor $\frac{\bar{N}}{N}$ scales the atom number $N$ of the single shot to the mean atom number $\bar{N}$ to avoid introducing correlations on all length scales.

From there, the $\langle\delta_c(x,y)\delta_c(x+\Delta_x,y+\Delta_y)\rangle$ correlation function is extracted from a central area within $75\,\%$ of the radius of the BEC. 
This correlator is a function only of the Euclidean distance $L = \sqrt{\Delta_x^2+\Delta_y^2}$, a reduction that is justified by the isotropy and homogeneity of the simulated space-time. 

The bare spectrum is calculated via a Hankel-Transformation to momentum space,
\begin{equation}
    \tilde{S}_k = \frac{2\pi m^2c_s^3}{\hbar k\lambda}\left(1+\frac{1}{2}k^2\xi^{2}\right)^{3/2}\int \mathrm{d}L L J_0(kL)\langle\delta_c\delta_c\rangle(L),
\end{equation}
where $J_0$ are Bessel functions of the first kind and order $0$, $m$ is the mass of $^{39}$K, $c_s$ is the extracted speed of sound in region III and $\lambda = \sqrt{8\pi\omega_z\hbar^3/m}a_{sc}$ is the interaction parameter depending on the transverse trap frequency $\omega_z$ and the scattering length $a_{sc}$.
The term in brackets stems from the Bogolibov dispersion relation. In the acoustic limit ($k\to 0$) the term reduces to one and the reported form in \cite{Viermann2022} is recovered.
The correlation function is only taken up to distances $L$ within $70\,\%$ of the maximal distance $1.5\,R$, where $R$ is the radius of the BEC, because for the largest distances the available statistics is much weaker. 
This is a limitation analog to the cosmic variance problem \cite{Dodelson2021}.

The experimental imaging system introduces a k-dependent sensitivity, due to a variety of optical abberations. This is corrected for by measuring the modulation transfer function of the imaging system $MTF(k)$, that captures the functional form of the $k$-dependence \cite{Chen2021,Hung2011}. To extract this, we drive the magnetic field multiple times close to the Feshbach resonance. As a consequence, most of the atoms are lost from the trap and the remaining ones form a low density gas. When taking images of this cloud, the individual atoms can be seen as point sources. We calculate a simple density contrast $\delta n(x,y) = n(x,y)/\bar n(x,y)-1$. Via Fourier transform to $\delta n(k_x,k_y)$ we get: $MTF(k_x,k_y) = |\delta n(k_x,k_y)|^2$, the two-dimensional modulation transfer function. Consequently, we subtract an offset, normalize the maximum to $1$ and take a radial average to reduce the result to one dimension. Finally, the result is smoothed, interpolated to the $k$ used in the Hankel transform and applied to the spectrum : $S_k = \tilde{S}_k/MTF(k)$.

To extract amplitudes and phases from the Sakharov oscillations, we fit $S_k(t)$ with $A_k\cos(2\omega_kt + \vartheta_k) + S_{k,0}$, where the fit's starting value for $\omega_k$ is set to the Bogoliubov frequency calculated from the measured $c_s$, and is restricted to be compatible with the standard error of the speed of sound. 
The offsets $S_{k,0}$ have start value $(\mathrm{max}_t(S_k(t))+\mathrm{min}_t(S_k(t))/2)$.
The amplitudes $A_k$ have start value $(\mathrm{max}_t(S_k(t))-\mathrm{min}_t(S_k(t))/2)$ and are restricted to positive values to avoid ambiguity in the phase.
Because for the linear expansions beginning and end are at a different scattering lengths, residual dynamics of the background density are observed even with adjusted trapping potential. This leads to an additional slow oscillatory modulation of the extracted spectra for all $k$. For the short hold times considered, this drift is well captured by a linear decrease of the spectrum in time and compensated for before the sinusoidal fit.

\subsection{Derivation of the theory curves}
\label{sec:DerivationTheoryCurves}

\subsubsection{General framework}
We study the dynamics of a free massless scalar field $\phi(t,r,\varphi)$ in an expanding (or contracting) flat $(2 + 1)$-dimensional spacetime determined by the FLRW metric or, equivalently in our analogy, the acoustic metric with vanishing background superfluid velocity, given in equation (1). This field $\phi$ is obtained by splitting the fundamental field $\Phi$ into a background field $\phi_0$ and two real scalar fields for the fluctuations: $\Phi = \phi_0 + \left(\phi_r  + i\phi\right)/\sqrt{2}$. In the acoustic limit, $\phi$ (which makes up the complex part of the fluctuation field) is connected to the real part $\phi_r$ via time derivative: $\phi_r \sim \partial_t\phi$. Because the gauge is chosen such that the background field is real, the real part is to first order responsible for the density fluctuations discussed before, such that: $\delta_c \sim \partial_t\phi$. The corresponding effective action for $\phi$ reads
\begin{equation}
    \Gamma_2 [\phi] = -\frac{\hbar^2}{2} \int  \text{d} t\, \text{d} r\, \text{d} \varphi \,\sqrt{g} \, g^{\mu\nu} \partial_\mu \phi \partial_\nu \phi,
    \label{eq:QEACurvedSpacetime}
\end{equation}
with the metric $g_{\mu\nu} = \text{diag}\left(-1,a^2(t),a^2(t) r^2\right)$, where $a(t) = 1/c_\mathrm{s}(t)$ and ${\sqrt{g}\equiv\sqrt{-\det(g_{\mu\nu})}}$. More details on this particular choice of fields can be found in \cite{Mireia2022}.

It is convenient to expand the field $\phi$ in terms of the time-dependent mode functions $v_k(t)$,
\begin{equation}
    \begin{split}
        \phi(t, \mathbf{r}) = \int \frac{\text{d}^2\mathbf{k}}{2\pi} \left[\hat{a}_{\mathbf{k}}  v_k (t) e^{-i\mathbf{k}\mathbf{r}} + \hat{a}^\dagger_{\mathbf{k}}  v_k^*(t) e^{i\mathbf{k}\mathbf{r}} \right],
    \end{split}
\end{equation}
where $\hat{a}^{(\dagger)}_{\mathbf{k}}$ are the (creation) annihilation operators in region I that follow the bosonic canonical commutation relations. The equation of motion for the scalar field $\phi$ is obtained from a variation of the action \ref{eq:QEACurvedSpacetime} with respect to the scalar field, leading the Klein-Gordon equation, and allowing us to write the mode equation given in (2).

As described in the main text and extensively discussed in \cite{Schmidt2024}, the components of the spectrum can be computed from the amplitudes of the analogue scattering problem. They follow from the matching conditions for the mode function $\psi_k = \sqrt{a} v_k$  at the beginning $\eta_\mathrm{i}$ and end $\eta_\mathrm{f}$ of region II which are given by 
\begin{equation}
\begin{aligned}
    \psi_k(\eta_\mathrm{i}) &= c_k\mathrm{e}^{- \mathrm{i} k  \eta_\mathrm{i}}, \\  \psi_k'(\eta_\mathrm{i}) &= - \mathrm{i} k  c_k\mathrm{e}^{-\mathrm{i} k  \eta_\mathrm{i}}  + \tfrac{1}{2}\dot a (\eta_\mathrm{i}) c_k \mathrm{e}^{-\mathrm{i} k\eta_\mathrm{i}} , 
\end{aligned}
\label{eq:MatchingLeft}
\end{equation}
and 
\begin{equation}
\begin{aligned}
    \psi_k(\eta_\mathrm{f}) &=  a_k e^{-\mathrm{i} k \eta_\mathrm{f}} + b_k e^{\mathrm{i}k\eta_\mathrm{f}},  \\
    \psi_k'(\eta_\mathrm{f}) &= -\mathrm{i} k [a_k e^{-\mathrm{i} k \eta_\mathrm{f}} - b_k e^{\mathrm{i} k \eta_\mathrm{f}}] \\
    &\qquad + \tfrac{1}{2} \dot a (\eta_\mathrm{f}) [a_k e^{-\mathrm{i} k \eta_\mathrm{f}} + b_k e^{\mathrm{i} k \eta_\mathrm{f} }],
\end{aligned}
\label{eq:MatchingRight}
\end{equation}
with the abrupt transitions from static to dynamic spacetime being taken into account by the terms proportional to the expansion (or contraction) rate $\dot a (\eta_\mathrm{i,f})$.

\subsubsection{Initial state}
We assume a thermal initial state, $\langle  \hat a_{k}^\dagger \hat a_{k'} \rangle = (1/2 +N _k^{\mathrm{in}} )  (2\pi) \delta(k-k')/k  $, where $N_k^{\mathrm{in}} = 1/\left[\exp\left( \frac{\hbar \omega_k}{k_B T} \right) - 1\right]$ is the Bose-Einstein distribution and $\omega_k$ is set by the Bogoliubov dispersion relation. 

\begin{figure}
    \centering
    \includegraphics[scale=0.35]{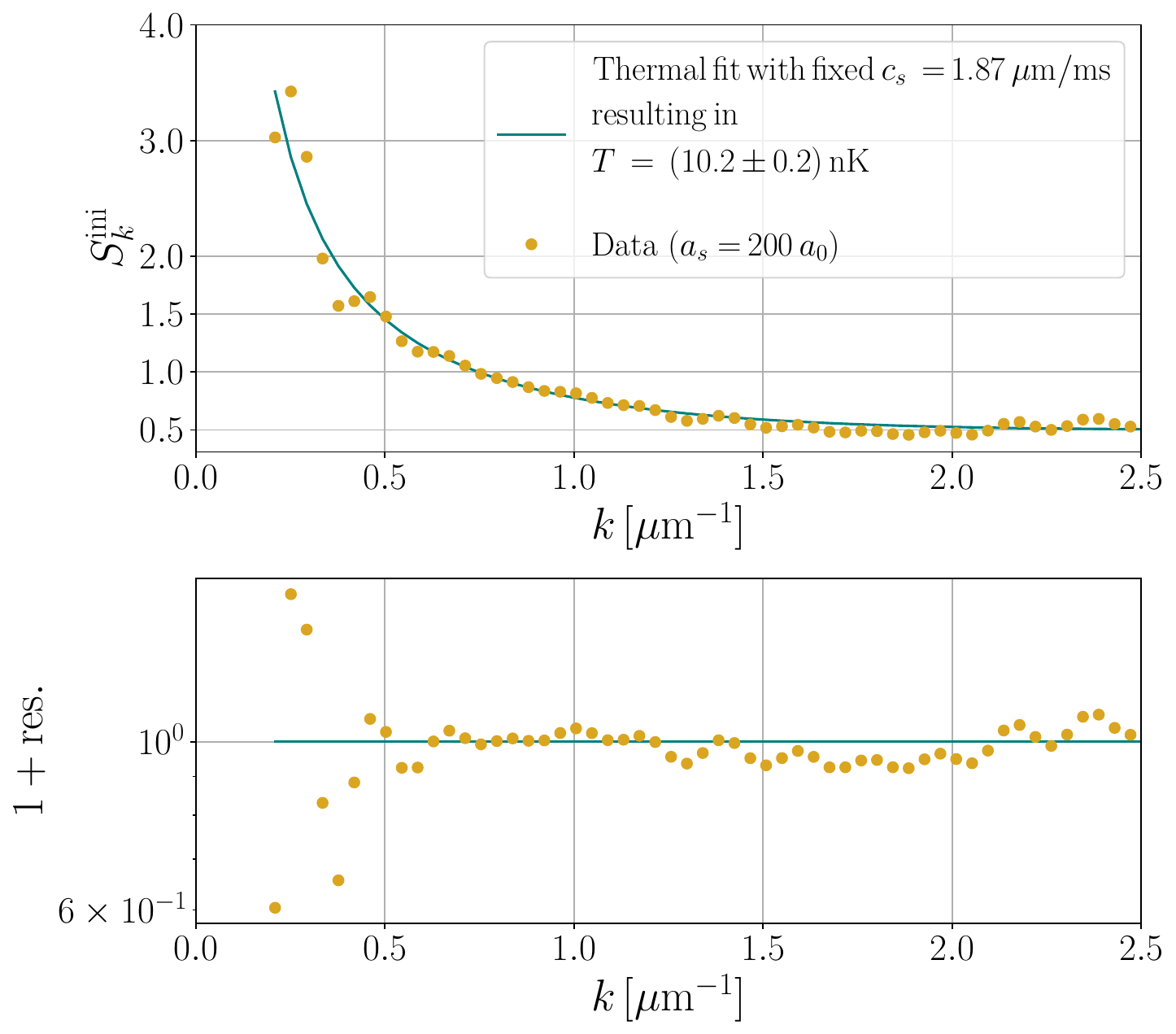}
    \caption{Initial spectrum of the harmonically modulated scale-factor scenario. The solid curve represents a thermal fit to the data points lying in the domain $0.2\, \mu\mathrm{m}^{-1} < k < 2.5 \mu\mathrm{m}^{-1} $. The lower panel shows the fit residuals centered around unity. }
    \label{fig:InitialSpectrumShaking}
\end{figure}

The temperature $T$ is determined by fitting a thermal distribution to experimentally extracted spectra of the state directly before dynamics, either at $400 a_0$ before the linear expansion or $200 a_0$ before the periodic dynamics. The independently extracted speed of sound is used, and the temperature is the only free fit parameter.
An example of this procedure is shown in Fig \,\ref{fig:InitialSpectrumShaking}. 
As described in further detail in \cite{Mireia2022}, the amplitude of the thermally stimulated particle spectrum is then 
\begin{equation}
    A_k = (1 + 2 N_k^\mathrm{in}) \abs{a_k b_k}/\abs{c_k}^2, \\
\end{equation}
whereas the phase remains insensitive to the initial population 
\begin{equation}
    \vartheta_k = \arg(-a_k^* b_k \mathrm{e}^{2\mathrm{i} k \Delta \eta}). 
\end{equation}

\subsubsection{Linear Expansion}
The scattering landscape analogue to the linear expansion, $a(t) = a_\mathrm{min}(1 + H_0 t) $ , can be described by flat box potential of height $H_0^2/4$ with $H_0 \approx 0.4 \, \mu\mathrm{m}^{-1}$ and width $\Delta \eta := \eta_\mathrm{f}-\eta_\mathrm{i} \approx \, 5.2 \mu\mathrm{m}$ and admits an analytic solution that is given by
\begin{equation}
 \begin{split}
     \frac{a_k}{c_k} =& \, \mathrm{e}^{\mathrm{i} k(\eta_\text{f}-\eta_\text{i})} \Big \lbrace \cos(\mu_k \Delta \eta)- \frac{k}{\mu_k} \sin(\mu_k \Delta \eta)\Big \rbrace ,\\
     \frac{b_k}{c_k} =& \frac{\mathrm{i}}{2} \frac{H_0}{\mu_k}  \mathrm{e}^{- \mathrm{i} k (\eta_\text{f}+\eta_\text{i})}  \sin(\mu_k \Delta \eta) 
 \end{split}
 \end{equation}
for subhorizon modes ($k > H_0/2$) with $\mu_k = \sqrt{k^2 - H_0^2/4}$. Further details can be found in \cite{Schmidt2024}. 

\subsubsection{Harmonically modulated scale-factor}
The harmonically modulated scale-factor of the shape $a(t)/a_\mathrm{min} = \tfrac{1}{2}[\gamma+1 + (\gamma-1)\cos(\nu t)] $ corresponds to the non-trivial potential 
\begin{equation}
\begin{aligned}
 V(\eta) &= \frac{\nu^2}{4} (\gamma-1) \gamma \\
 &\times \frac{ 1-\gamma + \left [ \gamma-1-\cos \left( \sqrt{\gamma}\, \nu \, \eta \right) \right ] \sec ^4(\sqrt{\gamma}\, \nu \,  \eta/2)}{\left [ 1 + \gamma \tan ^2 \left( \sqrt{\gamma}\, \nu \,  \eta/2 \right) \right]^2},
\end{aligned}
\end{equation}
where the scenario discussed in the main text can be captured by setting $\gamma = \sqrt{2}$ and $\nu \approx 5.9 \, \mu\mathrm{m}^{-1}$. In that case, the analogue Schrödinger equation is solved numerically in the dynamical region II and matched to plane-waves according to eqs.\@ \eqref{eq:MatchingLeft} and \eqref{eq:MatchingRight} with $\dot a (\eta_\mathrm{i,f})=0$.  

\subsubsection{Bouncing cusps}
For the cusp-modulation, the procedure has to be adjusted to the presence of the fast cusps in the scale factor and periods of static spacetime in between. Here it is convenient to adopt the transfer-matrix-method as described in \cite{Schmidt2024}. Let the transfer of the wavefunction along a single cusp be described by 
\begin{equation}
   \mqty(a_k^\mathrm{cusp} \, e^{-\mathrm{i}k  (\Delta \eta_\mathrm{cusp} + \eta_\mathrm{i})} \\ b_k^\mathrm{cusp} \, e^{\mathrm{i} k (\Delta \eta_\mathrm{cusp} - \eta_\mathrm{i}) } ) =  T_\mathrm{cusp} \mqty(c_k \, e^{-\mathrm{i}k\eta_\mathrm{i}}\\0)
\end{equation}
with 
\begin{equation}
T_\mathrm{cusp} = \mqty( (a_k^\mathrm{cusp}/c_k) \, \mathrm{e}^{-\mathrm{i} k \Delta \eta_\mathrm{cusp}} &  (b_k^\mathrm{cusp}/c_k)^* \, \mathrm{e}^{-\mathrm{i} k \Delta \eta_\mathrm{cusp}} \\ (b_k^\mathrm{cusp}/c_k) \, \mathrm{e}^{\mathrm{i} k \Delta \eta_\mathrm{cusp}} & (a_k^\mathrm{cusp}/c_k)^* \, \mathrm{e}^{\mathrm{i} k\Delta \eta_\mathrm{cusp}} ),   
\end{equation}
where $\Delta \eta_\mathrm{cusp}$ is the conformal time duration of the cusp.
Here, the $0$ in the initial state indicates no contribution with negative frequency, whereas the off-diagonal terms of $T_\mathrm{cusp}$ mix the positive and negative contributions.
The coefficients $a_k^\mathrm{cusp}$ and $b_k^\mathrm{cusp}$ are obtained via the matching conditions \eqref{eq:MatchingRight} evaluated at the end of the cusp where $\eta = \eta_\mathrm{i} + \Delta \eta_\mathrm{cusp}$ and are equivalent to $a_k$ and $b_k$, if region II consists of a single cusp. 
To calculate their values, the wavefunction $\psi_k$ to be matched is inferred from a numerical integration of the Schrödinger equation within a single cusp. First, the integration is carried out until $\eta = \eta_\mathrm{i} + \Delta \eta_\mathrm{cusp}/2$ where a delta-peak is situated. At this location, we analytically take into account that, according to similar reasoning entering Eqs.\@ \eqref{eq:MatchingLeft} and \eqref{eq:MatchingRight}, $\psi_k'$ changes discontinuously via $\psi_k' \to \psi_k' + \Omega \psi_k(\Delta \eta_\mathrm{cusp}/2)$ where $\Omega = \frac{1}{4} \,  \omega a_\mathrm{i}/\sqrt{2}$ is the sum of the contraction speed into the cusp and expansion speed away from the cusp. Then, the numerical integration is continued to $\eta = \eta_\mathrm{i} + \Delta \eta_\mathrm{cusp}$ where the transfer matrix elements are computed according to Eq.\@ \eqref{eq:MatchingRight} with $\dot a ( \eta_\mathrm{i} + \Delta \eta_\mathrm{cusp})=0$. The scenario in the main text corresponds to  $a_\mathrm{i} = 1/c_\mathrm{s}(200a_0)$ and $\omega = 2\pi \times 800 \mathrm{Hz} $ such that $\Omega \approx 0.4 \, \mu\mathrm{m}^{-1}$.

If region II contains multiple cusps, a holding period is included in between. The corresponding transfer matrix is $T_\mathrm{hold}(\Delta \eta_\mathrm{hold}) = \mathrm{diag}(\exp{- \mathrm{i} k \Delta \eta_\mathrm{hold}},\exp{\mathrm{i} k \Delta \eta_\mathrm{hold}})$ where $\Delta \eta_\mathrm{hold} = c_\mathrm{s}(200a_0) \Delta t_\mathrm{hold}$ is the sound-horizon during the period of constant scale factor and $\Delta t_\mathrm{hold} = 2 \mathrm{ms}$. 
With the given transfer matrices, the scenarios shown in Fig. 4\,b) can be captured by
\begin{equation}
\begin{aligned}
    &T_\mathrm{N-Cusps} = T_\mathrm{hold}(-\Delta \eta_\mathrm{lag}) \times T_\mathrm{hold}\left( - \tfrac{\Delta \eta_\mathrm{hold}}{2}\right) \\
    &\hspace{2cm}\times  (T_\mathrm{cycle})^N \times T_\mathrm{hold}\left( - \tfrac{\Delta \eta_\mathrm{hold}}{2}\right)
\end{aligned}
\end{equation}
where a single cycle can be described by $T_\mathrm{cycle} = T_\mathrm{hold}(\Delta \eta_\mathrm{hold}/2) \times T_\mathrm{cusp} \times T_\mathrm{hold}(\Delta \eta_\mathrm{hold}/2) $
and $T_\mathrm{hold}(-\Delta \eta_\mathrm{lag})$ accounts for a lag of the magnetic field with $\Delta t_\mathrm{lag} = 0.23 \, \mathrm{ms}$.

\subsection{Bogoliubov corrections in the scattering framework}
The dispersion relation of Bogoliubov excitations becomes linear only for low wavelengths. In particular the phase velocity 
\begin{equation}
   c_{\mathrm{ph}}= \frac{\omega_k}{k} = c_s \sqrt{1 + (k \xi)^2/2}
\end{equation}
approaches $c_s$ for small momenta, $k \ll \xi^{-1}$, where $\xi$ is the healing length, the inverse scale at which the free momentum term becomes relevant over the chemical potential $\mu = mc_s^2$.
However, even in the dispersive regime one can identify a momentum dependent cosmological scale-factor 
\begin{equation}
    a_k(t) \equiv \frac{1}{c_{\mathrm{ph}}(t)} = \frac{a(t)}{\sqrt{1 + \frac{1}{2} k^2 \xi^2(t)}},
\end{equation}
where $a(t)$ is the scale factor as defined in the main text and recovered for small momenta.
With this the mode equation can again be written in the form 
\begin{equation}
    \ddot v_k(t) + 2 \frac{\dot a_k(t)}{a_k(t)} \dot v_k(t) + \frac{k^2}{a_k^2(t)} v_k(t) = 0.
\end{equation}
Employing the rescaling
\begin{equation}
    v_k(t) = a^{-1/2}(t)\sqrt{1 + \tfrac{1}{2} k^2 \xi^2(t)} \psi_k(t),
\end{equation}
one finds the mode equation 
\begin{equation}
    - \psi_k''(\eta) + V_k(\eta) \psi_k(\eta) = 0
\end{equation}
in conformal time $\mathrm{d}\eta = \frac{1}{a(t)} \mathrm{d}t$ with the momentum dependent scattering potential 
\begin{widetext}
\begin{equation}
    V_k(\eta) =  \frac{\xi^{-4}(\eta) - 5 k^2 \xi^{-2}(\eta) + 4 k^4}{4 (\xi^{-2}(\eta)+k^2/2)^2} \dot a^2(\eta) +  \frac{1}{2} \frac{\xi^{-2}(\eta) - k^2/2}{\xi^{-2}(\eta) + k^2/2}  \ddot a(\eta) a(\eta)- \frac{1}{2} k^4\xi^{2}(\eta).
\end{equation}
\end{widetext}
This type of dispersive modification of the analogue cosmological spacetime below the healing length $\xi$ which takes the role of an analogue Planck-length has been studied for example in \cite{Weinfurtner2009}. 

For momenta much lower than $\xi^{-1}$ the potential in Eq.(4) is recovered. For higher k, the potential form differs, as the $\dot a$ term scales different from the $\ddot a$ term. Additionally a third term arises, that accounts for the dispersion and does not vanish in case of a static spacetime. 
The general framework described above can be still employed in the dispersive case where the solutions in the static regions I and III are now plane-waves that obey the Bogoliubov dispersion relation. 
However, the matching conditions in Eqs.\@ \eqref{eq:MatchingLeft} and \eqref{eq:MatchingRight} have to be generalized.

\bibliography{Bibliography}

\end{document}